\documentclass[onecolumn,11pt]{article}
\usepackage{times}
\usepackage{graphicx}
\usepackage[margin=1.0in]{geometry}

\usepackage[utf8]{inputenc}
\usepackage{enumitem,amssymb}
\usepackage{ragged2e}
\newlist{thematic}{itemize}{8}
\setlist[thematic]{label=$\square$}
\usepackage{pifont}

\usepackage[square,numbers]{natbib}

\usepackage{graphicx} 
\usepackage{longtable} 
\usepackage{rotating}
\usepackage{tabularx}
\usepackage{hyperref}
\usepackage{easylist}
\usepackage{amsmath}
\usepackage{bm}
\usepackage{array} % for extrarowheight
\usepackage{amssymb,amsmath} 
\usepackage{sidecap}
\usepackage{newunicodechar}
\DeclareRobustCommand{\okina}{%
  \raisebox{\dimexpr\fontcharht\font`A-\height}{%
    \scalebox{0.8}{`}%
  }%
}

\newcommand{\msun}{\ensuremath{\,M_\odot}}

\newcommand{\muHz}{\mbox{$\mu$Hz}}

\newcommand{\teff}{\mbox{$T_{\rm eff}$}}

\usepackage{titlesec}
\titlespacing{\section}{0pt}{3ex}{1ex}

\titlespacing{\section}{0pt}{2ex}{0.5ex}
\titlespacing{\subsection}{0pt}{2ex}{0.0ex}
\titlespacing{\subsubsection}{0pt}{2ex}{0.0ex}
\usepackage{natbibspacing}
\titleformat*{\section}{\large\bfseries}
\titleformat*{\subsection}{\normalsize\bfseries}

\begin{document}
\raggedright
\Large{\textbf{Roman CCS White Paper}} \linebreak

\large{\textbf{Asteroseismology with the Roman Galactic Bulge Time-Domain Survey}} \linebreak

\normalsize

\noindent \textbf{Scientific Categories:} Stellar physics and stellar types; Stellar populations and the interstellar medium \linebreak
  
\textbf{Principal Authors:}

Daniel Huber, University of Hawai{\okina}i (huberd@hawaii.edu) \& University of Sydney
 \linebreak
Marc Pinsonneault, Ohio State University (pinsonneault.1@osu.edu)
 \linebreak
%Phone: 808-956-8573 
% \linebreak
 
\textbf{List of contributing authors :}
\linebreak%(names and institutions)
Paul Beck, Universidad de La Laguna \& Instituto de Astrofisica de Canarias \linebreak 
Timothy R.\ Bedding, University of Sydney \linebreak
Joss Bland-Hawthorn, University of Sydney \linebreak
Sylvain N.\ Breton, INAF, Osservatorio Astrofisico di Catania \linebreak
Lisa Bugnet, Institute of Science and Technology Austria ISTA \linebreak
William J.\ Chaplin, University of Birmingham \linebreak
Rafael A. Garc\'\i a, Universit\'e Paris-Saclay, Universit\'e Paris Cit\'e, CEA, CNRS, AIM \linebreak %, 91191, Gif-sur-Yvette, France
Samuel K. Grunblatt, Johns Hopkins University \linebreak
Joyce A. Guzik, Los Alamos National Laboratory \linebreak 
Saskia Hekker, Heidelberg University \& Heidelberg Institute for Theoretical Studies (HITS) \linebreak
Steven D. Kawaler, Iowa State University \linebreak
St\'ephane Mathis, Universit\'e Paris-Saclay, Universit\'e Paris Cit\'e, CEA, CNRS, AIM \linebreak
Savita Mathur, Instituto de Astrof\'isica de Canarias \& Universidad de La Laguna \linebreak
Travis Metcalfe, White Dwarf Research Corporation \linebreak
Benoit Mosser, LESIA, Observatoire de Paris, Universit\'e PSL, CNRS, Sorbonne Universit\'e, Universit\'e de Paris  \linebreak
Melissa K. Ness, Columbia University \& Center for Computational Astrophysics \linebreak
Anthony L. Piro, Carnegie Observatories\linebreak
Aldo Serenelli, Instituto de Ciencias del Espacio (ICE, CSIC) \linebreak
Sanjib Sharma, Space Telescope Science Institute \linebreak
David R. Soderblom, Space Telescope Science Institute \linebreak
Keivan G.\ Stassun, Vanderbilt University \linebreak
Dennis Stello, University of New South Wales \linebreak
Jamie Tayar, University of Florida \linebreak
Gerard T. van Belle, Lowell Observatory \linebreak
Joel C. Zinn, California State University, Long Beach \linebreak

\newpage
\textbf{Abstract:}
Asteroseismology has transformed stellar astrophysics. Red giant asteroseismology is a prime example, with oscillation periods and amplitudes that are readily detectable with time-domain space-based telescopes. These oscillations can be used to infer masses, ages and radii for large numbers of stars, providing unique constraints on stellar populations in our galaxy.
%characterizing bulge stellar populations and interpreting microlensing planet demographics. 
The cadence, duration, and spatial resolution of the Roman galactic bulge time-domain survey (GBTDS) are well-suited for asteroseismology and will probe an important population not studied by prior missions. We identify photometric precision as a key requirement for realizing the potential of asteroseismology with Roman. A precision of 1\,mmag per 15-min cadence or better for saturated stars will enable detections of the populous red clump star population in the Galactic bulge. If the survey efficiency is better than expected, we argue for repeat observations of the same fields to improve photometric precision, or covering additional fields to expand the stellar population reach if the photometric precision for saturated stars is better than 1\,mmag. Asteroseismology is relatively insensitive to the timing of the observations during the mission, and the prime red clump targets can be observed in a single 70 day campaign in any given field. Complementary stellar characterization, particularly astrometry tied to the Gaia system, will also dramatically expand the diagnostic power of asteroseismology. We also highlight synergies to Roman GBTDS exoplanet science using transits and microlensing.

%\vspace{1cm}
%Instructions:
%2-3 pages of text, plus figures and references
%\url{https://roman.gsfc.nasa.gov/science/ccs_white_papers.html}
%Accordingly, Roman CCS white papers should include discussion of each of the following four elements: (1) the importance of the science investigation with respect to addressing open questions in its field, (2) how the capabilities of a Roman Core Community Survey will uniquely enable the investigation, (3) the minimal and optimal observational strategies for enabling the science investigation, within the bounds imposed on the relevant Core Community Survey by its science requirements in cosmology or exoplanet demographics, and (4) the impact of different observational strategy choices on the science investigation, expressed via appropriate metrics or figures of merit that the community survey committees can readily utilize. White papers discussing science investigations in any area of astrophysics, including cosmology and exoplanets, are encouraged, regardless of which Core Community Survey will enable them.

\justifying

\noindent

\pagebreak

\section{Background \& Motivation}

\noindent
Asteroseismology -- the study of stellar oscillations -- has been revolutionized by space-based time domain surveys such as CoRoT, Kepler/K2, and TESS. It has led to major breakthroughs in stellar astrophysics such as the discovery of rapidly rotating cores and magnetic fields in evolved stars \citep{beck12,deheuvels12,stello16,li22} and the systematic measurement of stellar masses, radii, and ages \citep{chaplin13a}. %Asteroseismology has also become the ``gold standard'' for calibrating indirect methods to determine stellar parameters such as surface gravity (\logg) from spectroscopy \citep{petigura17b} and granulation \citep{bastien13,kallinger16,bugnet2018,pande2018}, and age from rotation periods \citep[gyrochronology, e.g.][]{vansaders16} or magnetic activity \citep{metcalfe19}. 
While the discussion here focuses on stochastically-excited solar-like oscillators, many of these processes can also be probed in classical pulsators such as OB stars, Cepheids, RR Lyrae and $\delta$\,Scuti stars \citep{molnar18,kolenberg18,kurtz22}, providing a window into not only stellar physics, but galactic astronomy and -- through contributions to understanding the distance ladder --  to cosmology.

Asteroseismology of red giants is a particularly powerful tool for studying stellar populations. This is because red giants oscillate with $\gtrsim$\,0.1\,mmag amplitudes and periods of hours to days, allowing detections with moderate cadence and photometric precision out to large distances. Global frequency properties can be used to infer masses, ages, and evolutionary states for large numbers of giants \citep{pinsonneault14,Elsworth19}. Kepler asteroseismology was used as a fundamental calibrator in spectroscopic surveys \citep{jonsson2020} and Gaia \citep{creevey22}, yielded age estimates of nearby thin and thick disc stars \citep{aguirre18}, and has been used to age-date mergers of dwarf satellites \citep{montalban21}. It also uncovered unexpected populations, such as massive stars – typically an indication of youth – that have high alpha-capture to iron – typically, an abundance pattern indicating old age \citep{martig15}. While the origin of this population is contested \citep{jofre23}, it provides an example of the value of asteroseismology for population studies. 

However, asteroseismology has so far sampled only a limited portion of the galaxy. The seminal Kepler field is relatively nearby, and red giants had a complex selection function \citep{sharma16}. 
%As a result, although Kepler data were precise and accurate, they did not sample important Galactic populations. 
The K2 mission covered populations along the ecliptic plane, but with a relatively modest total sample size and depth \citep{stello17,zinn22}. TESS provides an all-sky survey that is large but shallow \citep{hon21} due to the small aperture. No space-based asteroseismic survey has so far sampled the crowded and distant stellar populations in the Galactic bulge. The bulge harbors a unique population born with high star formation efficiency in the earliest phases of Galactic evolution \citep{rix22}. Much of what we do not know about the formation of the Milky Way stems from our current inability to probe stellar populations in the inner Galaxy, which harbors a non axisymmetric bar whose chemodynamical evolution is complicated and poorly understood. Exploring how the  structure, kinematics and chemistry of the inner Galaxy depends on age offers a way to unlock its formation history. The bulge is also the closest analog to the spheroidal populations of other disk galaxies and entire elliptical galaxies, and thus will aid in understanding star formation in high-redshift galaxies currently studied with JWST. \textbf{The Roman Galactic Bulge Time Domain Survey (GBTDS) provides the first opportunity to apply the powerful tool of asteroseismology to measure masses and ages of stellar populations in the galactic bulge.} It will also serve as an important precursor for planned dedicated space-based asteroseismology missions in dense stellar environments \citep{haydn}.

\begin{figure}
\begin{center}
%\resizebox{11cm}{!}{\includegraphics{discomp_v3_1.png}}
%\resizebox{11cm}{!}{\includegraphics{discomp_v3_2.png}}
%\resizebox{11cm}{!}{\includegraphics{discomp_v3_3.png}}
\resizebox{\hsize}{!}{\includegraphics{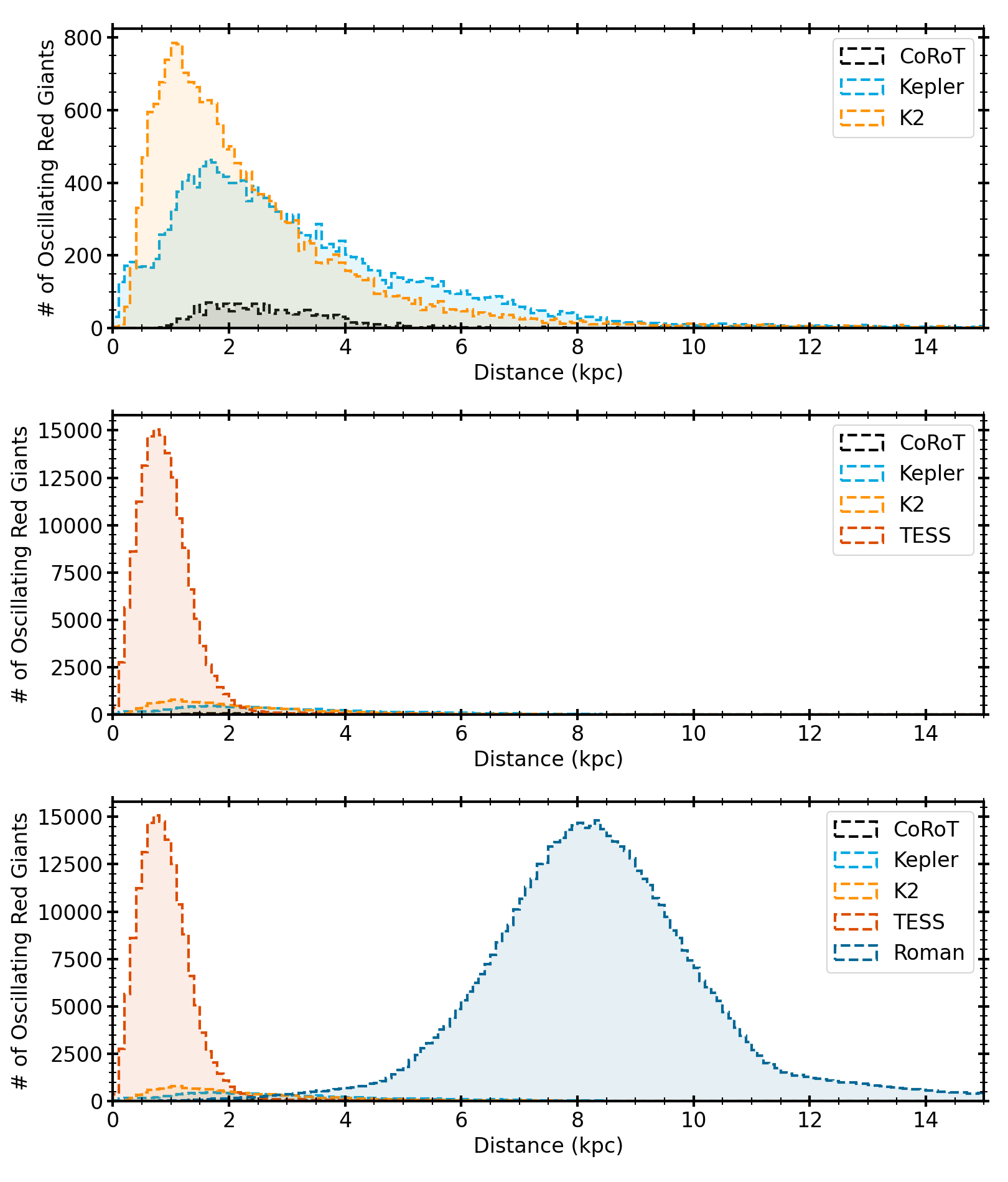}}
\caption{Distances of red giants with detected solar-like oscillations from space-based telescopes. Top panel: asteroseismic red giant yield from CoRoT ($\approx$\,2000 stars) \citep{hekker09}, Kepler ($\approx$\,16000 stars) \citep{yu18} and K2 ($\approx$\,20000 stars) \citep{zinn22}. Middle panel: same as top panel but adding the first all-sky asteroseismic yield from TESS (160,000 stars) \citep{hon21}. Bottom panel: same as middle panel but adding the expected yield from the Roman GBTDS ($\approx$ 600,000 stars) assuming photometric noise performance from \citet{gould15}.}
\label{fig:dis}
\end{center}
\end{figure}

Figure \ref{fig:dis} illustrates the potential of the Roman GBTDS for red giant asteroseismology. The Roman yield was estimated with a Galaxia simulation \citep{sharma14} of a 2.8 square degree FOV (corresponding to ten pointings) centered at (l,b)=(0.5,-1.5), using a photometric precision ($\sigma$) model for saturated stars \citep{gould15}, scaling relations for oscillation amplitudes ($A$) \citep{huber11,mosser11c} and requiring $\rm{SNR}=A/(\sigma/\sqrt{N})>15$ with $N=41472$ for six 70-day long campaigns with 15 minute cadence. Roman will for the first time perform space-based asteroseismology towards and beyond the galactic center, and is expected to increase the current yield by at least one order of magnitude ($\approx\,6 \times 10^{5}$ detections).

\section{Science Requirements}

\subsection{Photometric Precision} 
\noindent
Oscillation amplitudes increase linearly with luminosity, ranging from a few parts per million in the Sun to a few percent at the tip of the red giant branch. Photometric precision is the primary driver for the feasibility of asteroseismology, and in most cases is more important than cadence for a fixed photon noise. %For example, the Kepler, K2 and TESS missions have observed thousands stars with cadences fast enough to sample the oscillations, but the yield was limited by the per-cadence photometric precision.

\begin{figure}
\begin{center}
\resizebox{\hsize}{!}{\includegraphics{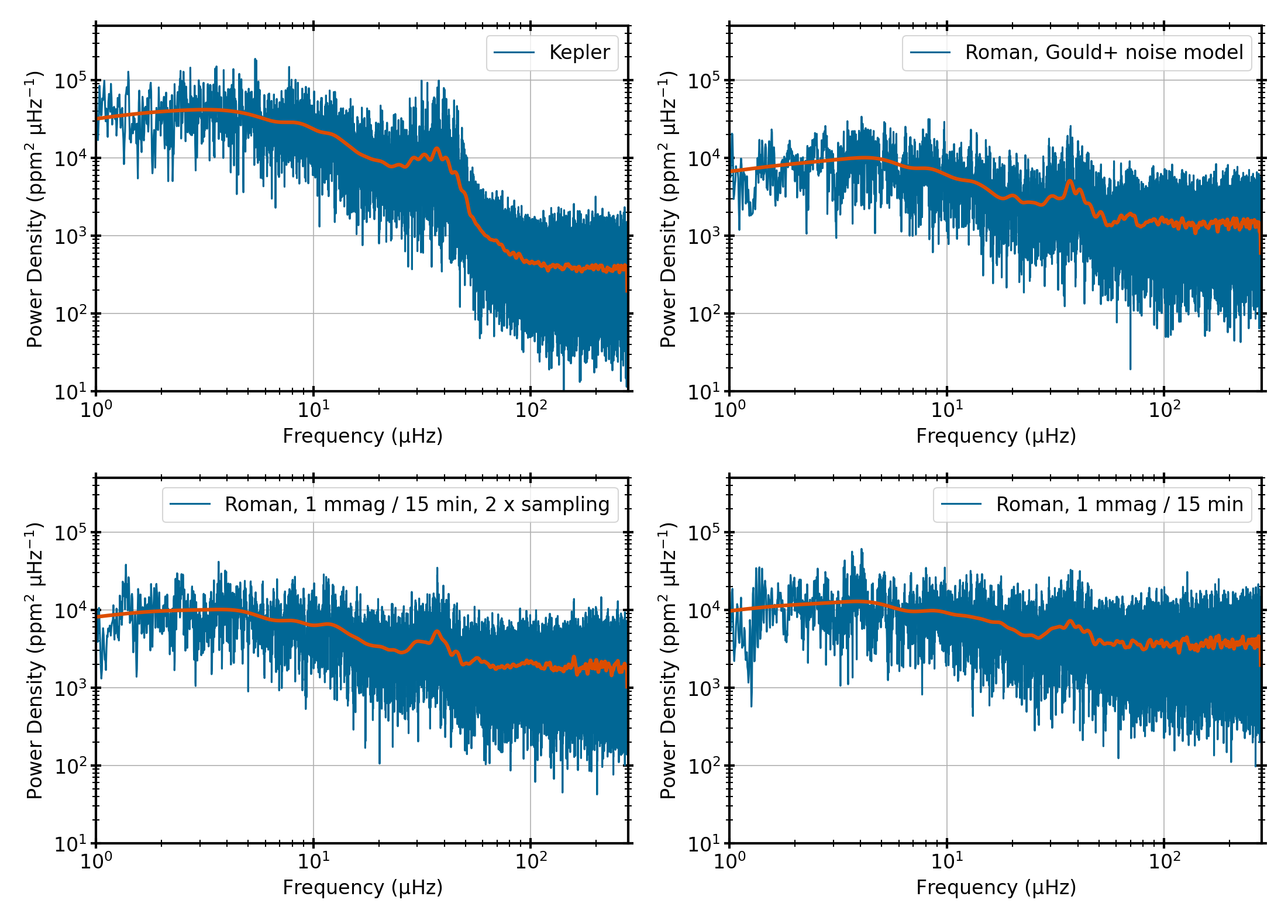}}
\caption{Power spectrum of the oscillating red clump star KIC\,2836038 observed by the Kepler space telescope using a nominal Roman GBTDS pointing strategy (six 70 day campaigns with 15 minute cadence, 3 campaigns at the start and 3 campaigns at the end of a 5-year period). Top left: original Kepler data (SNR$>300$). The power excess at $30-40$\muHz\ is due to solar-like oscillations. Top right: Simulated Roman power spectrum assuming the photometric noise performance following \citet{gould15} (SNR\,$\approx 15$). Bottom left: Same as top right but assuming nominal saturation noise (1\,mmag/15-min) but a twice faster sampling strategy (SNR\,$\approx 12$). Bottom right: Same as top right but assuming nominal saturation noise (1\,mmag/15-min) and cadence (SNR$\approx 9$). Power spectra were smoothed with a Gaussian filter with a width of 0.001\,\muHz\ (blue) and 1\,\muHz\ (red). The total expected yield for each noise scenario using a full synthetic stellar population (see bottom panel of Figure \ref{fig:dis}) is $6\times10^{5}$ stars (top right and bottom left panel) and $2\times10^{5}$ stars (bottom right panel).}
\label{fig:ps}
\end{center}
\end{figure}

A large fraction of red giants in a given stellar population are Helium-core burning (``red clump'') stars, a long-lived phase of stellar evolution following the tip of the red giant branch. A red clump star in the galactic bulge has $H\approx$\,13\,mag and thus will saturate the Roman detector in a single read. Techniques developed for saturated star photometry with Kepler/K2 \citep{white17} will not be directly applicable to Roman due to the different behaviour of H4RG detectors. Photometric precision estimates for saturated stars with Roman either assume a nominal noise floor of 1\,mmag / 15-min cadence \citep{montet17,penny19,wilson23}, or that the precision improves for brighter stars with careful modeling of the wings of the PSF \citep{gould15}. Figure \ref{fig:ps} shows a simulated red clump star power spectrum using the original Kepler data and different assumptions for photometric precision. Oscillations are clearly detectable ($\rm{SNR}\gtrsim 15$) with the improved saturated star precision, but become nearly undetectable ($\rm{SNR} < 10$) with a nominal noise of 1\,mmag/15-min, which would reduce the yield by a factor of 3. Assuming a strategy where Roman can observe fields twice as fast (and thus increase the nominal precision by a factor $\sqrt{2}$) nearly compensates for the SNR loss and would result in a similar yield as with improved saturated star precision, partially because of additional detections for non-saturated stars.  %\textbf{TODO: quantify this, e.g. through yield numbers.}

\textbf{We conclude that asteroseismology of red giant stars using the GBTDS requires a minimum photometric precision of 1\,mmag / 15-min or better for saturated stars.} Investigations of saturated star photometry will be critical for the success of GBTS asteroseismology.

\noindent
\subsection{Observing Cadence, Field Selection, and Filters} 
\noindent
Oscillation periods scale inversely with surface gravity, ranging from 5 minutes in Sun-like stars to weeks and months for evolved red giants. The nominal 15\,min GBTDS cadence yields a Nyquist frequency of 560\,\muHz, which 
%corresponds to oscillations in late subgiant stars and is thus sufficient given that the photometric precision limits detections to 
is sufficient for stars in and above the red clump as well as most other pulsators, except compact pulsators (e.g. hot subdwarfs and white dwarfs) and rapidly oscillating Ap stars.
%white dwarfs and rapidly oscillating Ap stars.%, which would require dedicated campaigns with 1-min(?) cadence.
%\textbf{Given the photometric precision, a faster cadence of the GBTDS will likely not significantly influence the asteroseismic science return if saturated star photometry can yield detections in red clump stars.} 
The upper cadence limit for red-giants is 30\,min, corresponding to a Nyquist frequency at the base of the red-giant branch.

\begin{figure}
\begin{center}
\resizebox{\hsize}{!}{\includegraphics{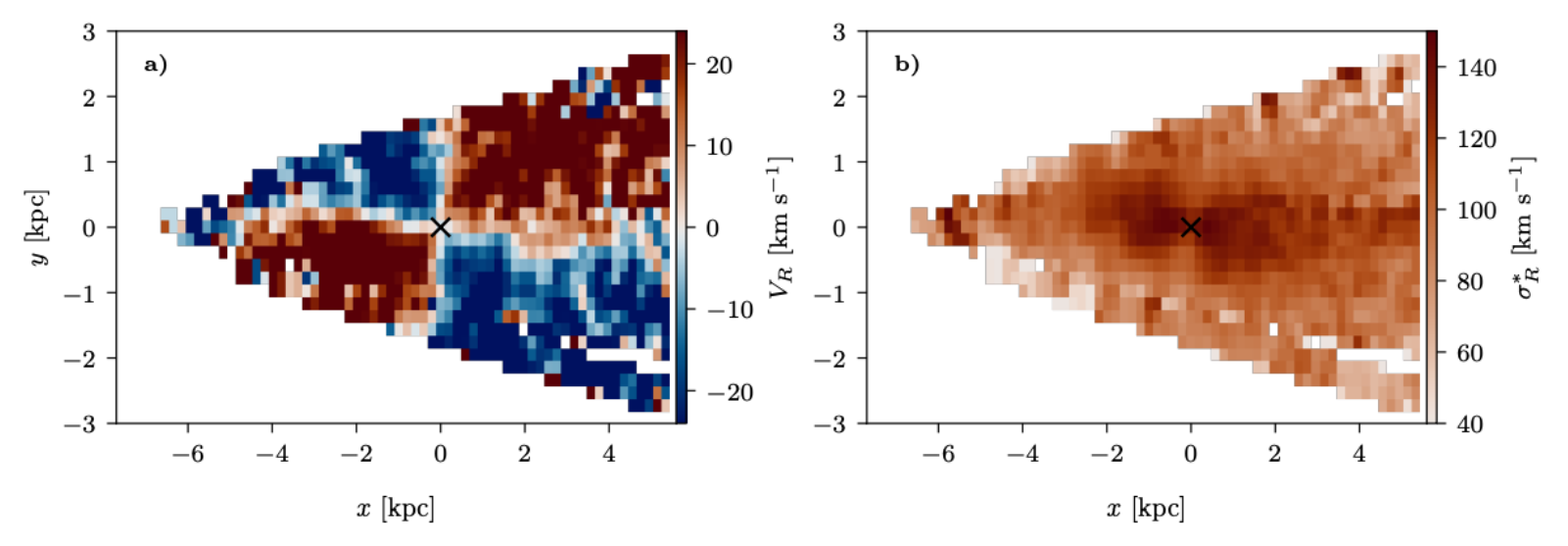}}
\caption{From \citet{hey23}: the first kinematic map (left: radial velocity; right: velocity dispersion) of the far side of the galactic bulge revealed by combining asteroseismic distances to luminous red giants observed by OGLE with Gaia DR3. The quadrupole pattern is the kinematic signature of the galactic bar. The Roman GBTDS will increase this sample in more extincted regions and provide masses, radii and ages that cannot be derived for the luminous giants for which oscillations can be detected from the ground.}
\label{fig:bulge}
\end{center}
\end{figure}

Asteroseismic inference is often divided into ``global asteroseismology'', which involves the determination of stellar radii, masses and ages, and ``boutique asteroseismology'', which aims to infer interior properties such as rotation. 
%The latter requires multiple years of continuous photometry to resolve rotational splittings of oscillation frequencies, and will in general not be feasible given Roman time baseline limitations. 
The latter will not be feasible with Roman given time baseline limitations, and require a dedicated asteroseismology mission in crowded fields such as HAYDN \citep{haydn}.
The former is sufficient to study stellar populations, and relies on sampling a large area to probe the formation history of our galaxy. Recent results using luminous giants from ground-based surveys have demonstrated the strong potential to study the galactic bulge, providing the first kinematic map of the far side of the galactic bar \citep{hey23} (Figure \ref{fig:bulge}). However, ground-based surveys only reach very luminous stars for which asteroseismic mass and age measurements are not possible. Roman will extend this success by enabling mass and age measurements over similar distance ranges (see Figure \ref{fig:dis}). \textbf{Increasing the number of fields observed by the Roman GBTDS would significantly enhance the asteroseismic science yield.} For example, additional fields covering a range longitudes will help explore the horizontal structure of the non-axisymmetric bar and fields at a wider range of latitudes would provide insights into the vertical structure of stellar populations in the bulge. 

Red giant asteroseismology is relatively insensitive to the timing of the GBTDS campaigns throughout the mission. However, pulsation timing for classical pulsators such as $\delta$\,Scuti stars \citep{murphy16,hey20} would benefit from having campaigns spaced as widely in time as possible.

%[additional ideas here? expand on this (DS: Maybe ask Sanjib???)]

%\noindent
%\subsection{Filters}
%\noindent
Oscillation amplitude increases for blue wavelengths \citep{KB95}.
%since oscillations can to first order be described as temperature fluctuations \citep{KB95}. 
Compared to Kepler, oscillation amplitudes decrease by $\approx$\,65\% in the Roman F149 filter \citep{gould15}. Detecting oscillations in multiple passbands can be used for mode identification, which is important for classical pulsators. Observing the same fields in multiple filters thus provides benefit for the time-domain study of classical pulsators ($>$1.3\msun), but overall has lower priority than increasing the number of fields.

\noindent
\subsection{Astrometry and Saturated Star Photometry} 
\noindent
Global asteroseismology collapses the oscillation spectrum into two properties: the frequency of maximum power $\nu_{\rm max}$ and the average frequency spacing $\Delta \nu$. When combined with $T_{\rm eff}$ and abundances, these data can be used to infer mass, radius, and age.  However, 
%it is easier to detect a power excess (measure $\nu_{\rm max}$)  than to characterize the frequency pattern (measure $\Delta \nu$) . 
we will be able to measure $\nu_{\rm max}$ for many more targets than those for which $\Delta \nu$ can be measured. 
Fortunately, there is an alternative: one can combine an independent radius with $\nu_{\rm max}$ to infer mass without $\Delta \nu$. Radii can be inferred from a combination of Gaia luminosity and $T_{\rm eff}$. The resulting mass uncertainties are competitive with those from full asteroseismic characterization \citep{stello22}. In the bulge, extinction map uncertainties in the optical will result in large errors in Gaia luminosities.
%even in cases where $T_{\rm eff}$ is measured. 
However, idifferential Roman astrometry could be used to infer precise relative parallaxes \citep{gould15}; when tied to the Gaia system, this could be translated to precise radii. Since red clump stars have similar intrinsic \teff, their radii can be inferred even without detailed spectroscopic data.

Red clump stars are saturated at the distance of the Galactic center, which could be used to centroid images using diffraction spikes \citep{gould15}. \textbf{Both the detection of oscillations and precise Roman parallaxes will therefore benefit from an emphasis on precise and accurate saturated star astrometry and photometry.}

%\begin{itemize}

%\item cadence shorter than 15 minutes to reach higher numax? requires a reduction in overall fields
%\item slower cadence to cover a larger area of the bulge?
%\item advocate for high-cadence visit for shorter duration to get less evolved stars?
%\item observe in multiple filters to probe amplitude excitation?

%\item Nominal strategy: 60-72 day seasons, 3 seasons at the beginning of the 5 year mission (2 at the end?). For each season observe 7 fields with exposure time of ~1 minute; with slewing will yield 15 minute cadence (7\% duty cycle).

%\item During a community update meeting, Julie McEnery mention slew times may be more efficient than expected. This could a) yield higher cadence / better duty cycle or b) be used to do something else. Tom Barclay mentioned idea of spending the first few days of an observing campaign observing only two pointings, but at higher cadence. e.g. you could do something like 2 min cadence for 5-10 days. 

%\item make a case for photometric precision. more luminous stars? better for high. don't need long timebaseline, better to add more fields. sidesteps problem with saturated star photometry. more targets with 70d dwell time or better photon statistics for the same targets.

%\end{itemize}

\section{Synergies with Exoplanet Science}

\subsection{Characterization of Stellar Populations for Transiting Exoplanet Demographics} 
\noindent
The Roman GBTDS will detect tens of thousands of transiting exoplanets \citep{montet17,wilson23}. 
%Similar to Kepler/K2 and TESS, the cadence, precision, and baseline requirements for transiting exoplanet science and asteroseismology are similar, with the exception that the majority of transiting planet hosts will not saturate. 
Because most asteroseismic detections will be in red clump stars, stars with detected oscillations and transiting planets will be rare. However, GBTDS asteroseismology will constrain the underlying stellar population in the galactic bulge, which will be important for interpreting the exoplanet yield. For example, the demographics of transiting exoplanet hosts (such as their distances and association to the thin and thick disc populations) will be sensitive to the importance of host star abundances on planet formation \citep{wilson23}. Asteroseismology will complement this by mapping the mass and age distributions of the bulge, thin disc, and thick disc populations in the same galactic regions where transiting exoplanets will be found.

\subsection{Characterization of Microlensing Lens and Source Stars}
\begin{figure}
\begin{center}
\resizebox{\hsize}{!}{\includegraphics{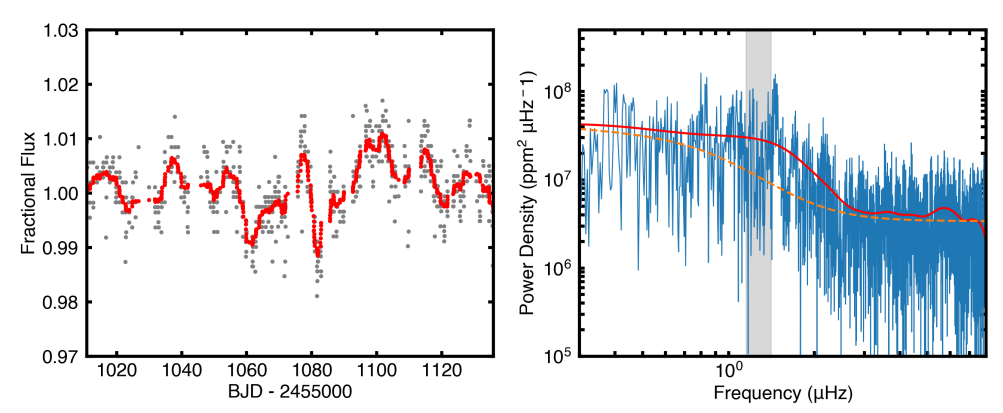}}
\caption{From \citet{li2019}: the first application of asteroseismology for a source star of a microlensing event. Left: portion of the OGLE light curve showing the oscillation signature of the source red giant. Right: Power spectrum showing the oscillation excess centered at $\approx$\,1\muHz. The red line is a heavily smoothed power spectrum and the orange dashed line is the background model. The oscillations were used to calculate an independent distance to the source star.}
\label{fig:lens}
\end{center}
\end{figure}

\noindent
Characterizing exoplanets using microlensing requires knowledge of distances to the lens and the source stars.
%to break degeneracies in the lensing geometry [add ref]. 
Distances to source stars are typically assumed since in most cases the lens and source star cannot be resolved. In some cases, however, the source star will be an evolved red giant, which will result in oscillation signal in the microlensing light curve. Stellar oscillations can be used to derive precise distances by constraining the luminosity, and thus help constrain the properties of planets detected using microlensing. First detections have already been made using OGLE (Figure \ref{fig:lens}) \citep{li2019}.
%, and many more [quantify?] can be expected with Roman. 
\textbf{Time-domain stellar variability from oscillation and granulation imprinted on microlensing light curves may allow the systematic characterization of lens and source stars in microlensing detections with the Roman GBTDS.}

\section{Conclusions}
\noindent
%Time domain missions lend themselves naturally to multiple scientific topics. 
There are strong links between the goals for exoplanet demographics and asteroseismology with the Roman GBTDS. The most natural linkage is the study of red clump (core He-burning) stars in the Galactic bulge. %The currently planned observing cadence and field selections are a good match to the science requirements for asteroseismology. 

Relative to microlensing studies, asteroseismology is more sensitive to photometric precision, and prime asteroseismic targets will likely be saturated at the distance of the Galactic center. As a result, controlling photometric noise at the level of 1\,mmag/15-min cadence or better is a key scientific requirement for Roman GBTDS asteroseismology. 
%A higher noise level will still permit some results, but with a much reduced sample size.
If saturated star photometry is limited to 1\,mmag/15-min cadence, observing the nominal fields at higher cadence will be important to reduce noise. If saturated star photometry performs better than 1\,mmag/15-min cadence, observing additional fields will increase the scientific yield by probing additional stellar populations. The minimum cadence requirement is 30 minutes.

A second synergy is precise and accurate astrometric data for saturated stars. %Because of the differences between IR and optical detectors, this is feasible to do. 
In particular, tying relative Roman astrometry to the Gaia system is a promising approach that could yield reliable mass estimates for low signal to noise asteroseismic detections. 
%It could also improve the overall distance model for microlensing studies.
The requirements for red giant asteroseismology with the Roman GBTDS will map to most other types of pulsators in the H-R diagram. Asteroseismology will also provide natural synergies with exoplanet science, including both a deeper understanding of the underlying stellar populations and the characterization of lens and source stars in microlensing events.

%Our third goal is stellar characterization; complementary ground-based follow-up and calibration of photometric diagnostics will be important for understanding bulge populations.

\newpage

\bibliographystyle{unsrtnat}
\bibliography{references}

\end{document}